\documentclass{ceurart}

\usepackage{float}
\usepackage[nameinlink]{cleveref}
\usepackage{lscape}
\usepackage{longtable}

\usepackage{listofitems}

\begin{document}

\copyrightyear{2025}
\copyrightclause{Copyright for this paper by its authors. Use permitted under Creative
Commons License Attribution 4.0 International (CC BY 4.0).}
\conference{KGCW'25: 6th International Workshop on Knowledge Graph Construction, June
1st, 2025, Portorož, SLO}

\author{Ademar {Crotti Junior}}[%
	orcid=0000-0003-1025-9262
]
\author{Christophe Debruyne}[%
	orcid=0000-0003-4734-3847,
	email=C.Debruyne@uliege.be,
]
\address{Montefiore Institute, University of Liège, Belgium}

\title{A Protocol for KG Construction Tasks Involving Users}

\begin{abstract}
Knowledge graph construction (KGC) from (semi-)structured data is challenging, and facilitating user involvement is an issue frequently brought up within this community. We cannot deny the progress we have made with respect to (declarative) knowledge graph construction languages and tools to help build such mappings. However, it is surprising that no two studies report on similar protocols. This heterogeneity does not allow for comparing KGC languages, techniques, and tools. This paper first analyses studies involving users to identify the points of comparison. These gaps include a lack of systematic consistency in task design, participant selection, and evaluation metrics. Moreover, there needs to be a systematic way of analyzing the data and reporting the findings, which is also lacking. We thus propose and introduce a user protocol for KGC designed to address this challenge. Where possible, we draw and take elements from the literature we deem fit for such a protocol. The protocol, as such, allows for the comparison of languages and techniques for the RDF Mapping Language (RML) core functionality, which is covered by most of the other state-of-the-art techniques and tools. We also propose how the protocol can be amended to compare extensions (of RML). This protocol provides an important step towards a more comparable evaluation of KGC user studies. 
\end{abstract}

\begin{keywords}
	KG Construction \sep User studies \sep Research Methods
\end{keywords}

\maketitle

\section{Introduction}
We preach to the choir that knowledge graphs are essential for meaningfully organizing and representing information in various domains. However, as knowledge graphs grow in complexity, efficient methods for their construction (or generation) are crucial. When dealing with the challenges of (semi-)structured data sources, such as the lack of explicit semantics, which need to be aligned with ontologies or vocabularies, creating such mappings becomes a \textit{knowledge engineering task} where \textit{user involvement is crucial}. Users bring the necessary domain expertise to ensure the mappings are appropriate. 

Scholars have systematically analyzed the functionalities of Knowledge Graph Construction (KGC) tools and proposed benchmarks to analyze their behavior in different settings and their memory and CPU usage. It is thus surprising that user involvement and the perception of the user using the languages, tools, etc., have yet to be studied in such detail. Conducting such a study for all languages and tools is an infeasible undertaking for one group, but what is feasible is putting forward a protocol that scholars in the domain should adopt to report on user studies. This paper aims to achieve this goal by proposing a user study protocol for KGC.

This will enable researchers to compare different knowledge graph construction languages and techniques, leading to a better understanding of their strengths and weaknesses and, ultimately, to more effective tools for KGC.

This paper's contributions are twofold. First, we review user studies in the KGC domain, which indicate the abovementioned challenges. The second contribution is the protocol. Section \ref{Related Work} provides an overview of the related work on (declarative) approaches to KGC by mapping data sources onto RDF datasets, focusing on those that explicitly report on user studies. The goal is to show that no two papers adopt the same protocol, which makes comparing studies impossible. Section \ref{Proposed User Study Protocol} presents the protocol we have made available with CC-BY-SA 4.0 license. The protocol provides detailed guidelines for recruiting participants and disclosing potential biases. The process guidelines for informed consent, pre-questionnaires, familiarization activities, task execution, and post-questionnaires. The tasks consist of five tasks, of which, when comparing two groups, the last two can be changed to ensure a common base for comparison. 
The related work will show that reporting is often limited to simple metrics and averages. Still, we deem it important to analyze the relationships between task execution, perceived usefulness, and perceived cognitive load. To this end, Section \ref{Results and Analysis} proposes the statistical means to use when adopting this protocol. In Section \ref{Discussion}, we discuss the resources from various aspects, such as the scientific and technical, to elaborate on the soundness of our approach. This section also discusses some of the limitations. Section \ref{Conclusions} then concludes the paper and proposes future directions.

\section{Related Work}
\label{Related Work}

In \cite{DBLP:journals/ws/AsscheDHHMD23}, the authors presented an excellent survey on declarative KGC tools to help the community and practitioners choose which languages, tools, or techniques fit their needs. However, the article looks at those from a technical perspective. They look at the functionalities offered by the different options. In \cite{DBLP:conf/semweb/AsscheCD24}, the authors proposed a benchmark to compare KGC tools and applied it to some well-known implementations such as RMLMapper\footnote{\url{https://github.com/RMLio/rmlmapper-java}}, Morph-KGC \cite{DBLP:journals/semweb/ArenasGuerreroCTPC24}, and SDM-RDFizer \cite{DBLP:conf/cikm/IglesiasJCCV20}. It is surprising to see that the perceptions of users and practitioners have yet to be examined in a systematic manner. 

From a broader perspective, \cite{10.1109/TVCG.2023.3326904} describes three ``personas'' that engage with KGs: KG builders, KG analysts, and KG consumers, which were distilled from interviews with practitioners. As the name intuitively implies, the KG builder persona would be responsible for generating the KG from heterogeneous sources, but the persona is also in charge of ontology engineering. The authors state that builders could benefit from tools that help them ensure that the schema is respected (what the authors call an ``enforcer'') as well as adequate visualization tools. While the paper does not explicitly mention KG construction and mappings as tasks, they fall under the ``data integration'' umbrella. The interviews indicate that there are challenges impeding uptake.

It seems that practitioners' or users' roles are sometimes neglected. This is certainly the case for KG construction, as we will demonstrate via our literature review. Our review looked at the following papers reporting on users, their experiences, and/or perceived usability: \cite{DBLP:conf/esws/PinkelBHMST14,DBLP:conf/esws/HeyvaertDHVSMW16,DBLP:journals/semweb/SiciliaNN17,DBLP:conf/semweb/BakBL17,DBLP:conf/esws/JuniorDO17,DBLP:journals/ws/HeyvaertDMSHVSM18,DBLP:conf/hworkload/JuniorDLO18,junior2019jigsaw,DBLP:journals/peerj-cs/Garcia-Gonzalez20,warren2024path,DBLP:journals/biomedsem/BrouwerBASDVTO24}. Most of these studies looked at the creation of mappings. Exceptions are \cite{junior2019jigsaw} reporting on studies on mapping understanding and \cite{DBLP:journals/biomedsem/BrouwerBASDVTO24} reporting on a complex data flow that included mapping creation.\footnote{Excluded from this survey are publications that do not report on users. For example, in \cite{DBLP:conf/esws/HeyvaertDVMW15}, the authors reported involving participants, but no report on the participants' experience was made. Other examples of papers mentioning users, participants, etc. without any detailed reporting include \cite{DBLP:journals/softx/IbrahimCSSHI24}, \cite{DBLP:conf/esws/HeyvaertMDV18}, and \cite{DBLP:conf/icwe/Iglesias-Molina23}.} We compare the various aspects of these user studies in Tables \ref{tab:table1-1}, \ref{tab:table1-2}, and \ref{tab:table1-3}. From these tables, we can observe a couple of important points:

\begin{itemize}
    \item Some report on comparing mapping languages and/or tools (e.g., \cite{DBLP:journals/ws/HeyvaertDMSHVSM18} and \cite{DBLP:conf/hworkload/JuniorDLO18}), and others report on comparing mapping languages (e.g., \cite{DBLP:journals/peerj-cs/Garcia-Gonzalez20} and \cite{warren2024path}). Quite a few papers merely report on the perceived usability of their tool without any comparison. We argue that reporting on user studies only makes sense if there is a basis for comparison. Without a reference point (e.g., a comparable tool evaluated under a shared protocol) or even a common protocol to establish such points, it becomes difficult to interpret the significance of usability results. For instance, reporting that users found a tool ``easy to use'' or completed a task within a certain time frame lacks context unless these outcomes can be meaningfully compared. Comparative user studies are therefore essential to provide this community with insights and to guide the development of more effective knowledge graph construction languages and tools.
    \item Looking at the procedure, we see many recurring elements (some (training) resources are being shared, pre-assessment surveys, introduction of tasks, surveys, etc.). No two procedures are the same, which limits our ability to compare the studies. Some studies reported asking about information such as gender and age but did not report on those in the data analysis.
    \item Most studies involved participants with expertise in IT, databases and/or semantic web technologies. Many studies also report inviting MSc students in computer science or related fields. Self-reported prior knowledge and competencies are a recurring theme, but no two studies tackle this aspect comparably. 
    \item The same heterogeneity can be observed for the tasks and datasets, where we do notice that most studies adopt datasets that do not require specific domain expertise (people, movies, places, etc.).
    \item Recurring themes in data being analyzed are time (efficiency), accuracy, and perceived usability. Most rely on System Usability Metrics \cite{10.5555/2817912.2817913} (SUS) for perceived usability. A few studies rely on Post-Study System Usability Questionnaire \cite{lewis2002psychometric} (PSSUQ) to obtain information on perceived information quality, interface quality, and system usefulness. Such studies do allow a means to compare results. Few studies have reported on qualitative feedback from users and the mental workload of tools and mapping languages.
    \item Most studies merely report on averages, which can arguably make sense when authors only report on one group and tools or languages are not compared. Few studies employed techniques to analyze whether groups are (significantly) different or whether certain aspects had a (statistically significant) impact on efficiency, accuracy, etc. 
\end{itemize}

From this survey, we can conclude that there is a critical need for homogeneous protocols, including tasks, for comparing advances in KG construction (KGC) approaches (mapping languages and tools alike). In the next section, we propose a protocol to address these issues and how they can be used.

\begin{table}[!h]
    \centering
    \caption{Comparison of KG construction languages, techniques, and tools in which participants were observed. We can observe that procedures vary widely.}
    \label{tab:table1-1}
    \includegraphics[width=\linewidth,page=1,trim={0 1.3in 0 0 0},clip]{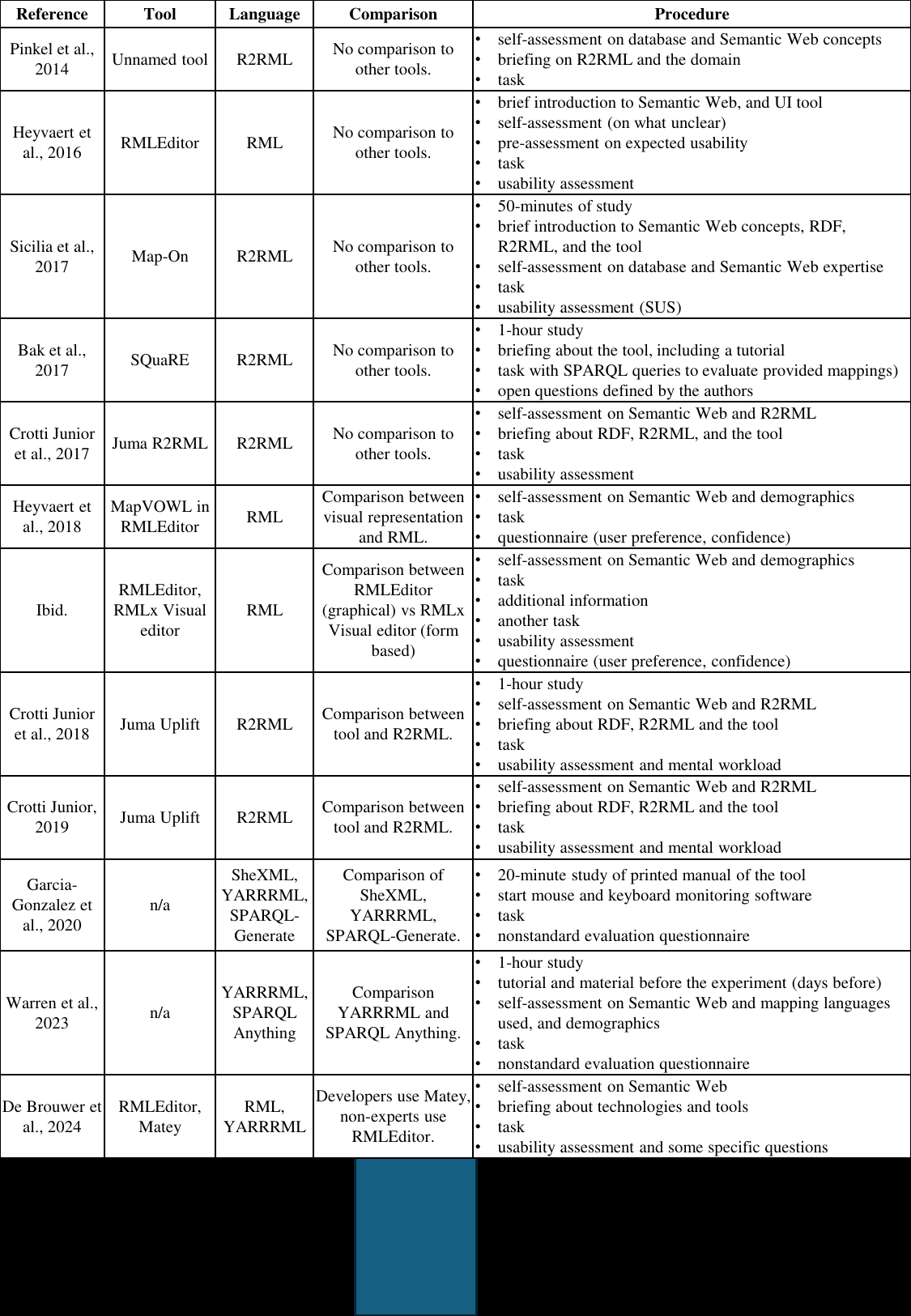}
\end{table}

\begin{table}[!h]
    \centering
    \caption{Comparison of KG construction languages, techniques, and tools in which participants were observed. We can observe that no two papers share tasks and data analysis techniques.}
    \label{tab:table1-2}
    \includegraphics[width=\linewidth,page=2,trim={0 1.63in 0 0 0},clip]{comparison.pdf}
\end{table}

\begin{table}[!h]
    \centering
    \caption{Comparison of KG construction languages, techniques, and tools in which participants were observed. We can observe that self-reported prior knowledge, self-reported expertise, and participant selection vary widely. Also, the number of participants varies.}
    \label{tab:table1-3}
    \includegraphics[width=\linewidth,page=3,trim={0 3.63in 0 0 0},clip]{comparison.pdf}
\end{table}

\section{The KG Construction User Study Protocol}
\label{Proposed User Study Protocol}

This section presents the protocol for comparing KGC tools and languages. The protocol's\footnote{\url{https://github.com/chrdebru/kgc-user-study-protocol}} structure foresees placeholders for text to be easily adapted for research ethics applications.

The protocol can be used to analyze the perceived usability and cognitive load of a mapping language or tool, as well as the accuracy achieved by users and the task execution time. When scholars use this protocol on only one group, the hypotheses are limited to comparing participants with different backgrounds or comparing the results with other experiments using that same protocol. Scholars adopting this protocol can easily extend the protocol to compare two tools, techniques, or languages. This will be explained in Section \ref{Variants}.

The focus of the protocol is to facilitate the discovery of problems (i.e., formative testing), not on task-level measurements. \cite{10.1145/1167948.1167973} describes the difference between the two. The protocol proposes task-level measurements and techniques for analyzing the data. When sample sizes are small, they can merely give insights.

\subsection{Participant Selection}
Adopters of the protocol should indicate how participants are recruited and from where. Adopters should disclose potential biases by providing details about factors influencing the study or participant behavior. Examples include the hierarchical relationship between research group leaders and researchers, as well as students recruited from classes. There is also a difference between voluntary participation and mandatory representation (e.g., in the context of a teaching activity), which may lead to non-probability samples. The absence of certain participants may lead to response bias, which is the possible impact on observations had those participants taken part in the experiment. In \cite{creswell2017research}, the authors describe how disclosing participant selection is important to recognize potential biases.

Practitioners involved with UI and UX often state that five users are sufficient to discover most of the usability problems. That is more likely the case for problem discovery than task-level measurements, which require larger sample sizes \cite{10.1145/1167948.1167973}. As \cite{faulkner2003beyond} observed in an experiment, \textit{``increasing the number from 5 to 10 can result in a dramatic improvement in data confidence.''} They also found that increasing the number to 20 practically guarantees all problems to be seen, but we recognize that recruiting participants is difficult. As such, we (strongly) recommend 10 participants per group.

\subsection{Process} Participants begin by reviewing informed consent materials and completing a pre-questionnaire assessing their backgrounds, prior knowledge, and expectations. Next, they attend a presentation introducing the technology, review relevant documentation, and engage in a familiarization activity. The core of the study involves participants executing a defined task using the technology. Finally, participants complete a post-questionnaire evaluating their experience, including usability, efficiency, and perceived cognitive load. This structured process aims to gather comprehensive data on user interaction and perception of the technology.

Next to presenting and demonstrating the tool or mapping language, we also request participants to handle the environment. We deem this familiarization activity novel compared to the related work. This activity ensures participants are comfortable executing mappings within the provided (tool's) environment. We guide participants in demonstrating the practical aspects of using the tool's interface, such as utilizing the command line in the terminal or identifying the correct buttons to click. This focused familiarization will prevent the environment from becoming an obstacle, allowing us to assess the tool or language's usability and gather unbiased feedback on its functionalities.

Furthermore, we ask authors to report on how responses were submitted (e.g., email, paper, form) and the anticipated duration of the experiment. While in-class experiments often have time constraints, other environments may be more flexible. Our protocol foresees 1 hour for the five tasks. If, for example, all steps are conducted in a classroom setting, the experiment would require 2:30 in total (i.e., including the questionnaires and consulting the training material). Finally, clarify whether participants can ask questions during the experiment. Help should be limited to aspects not core to the KG construction process and experiment. For example, helping participants navigate to the correct directory in a terminal or assessing whether a network issue is permitted, but providing help to execute a mapping is not. Ideally, studies should report on those (and their number of occurrences).

\subsection{Pre-questionnaire}
Studies often inquire about participants and group them based on self-reported information on their background and proficiency with specific techniques. We aim to homogenize this by proposing an exhaustive list of current roles, formal training, and self-perceived competency levels in certain Semantic Web technologies. We also included three questions related to intrinsic motivation (enjoyment, curiosity, and value). It is important to analyze the impact of self-selection bias in voluntary participation. This allows us to explore potential self-selection bias in voluntary participation by examining correlations between motivation levels, task performance, and perceived usability. In contrast to self-selection bias in voluntary participation, mandatory settings introduce the risk of low engagement among less intrinsically motivated participants. Including questions on intrinsic motivation enables us to assess how motivation levels influence task performance and perceived usability.

\subsection{Mapping tasks}
\begin{figure}
    \centering
    \includegraphics[width=0.9\linewidth]{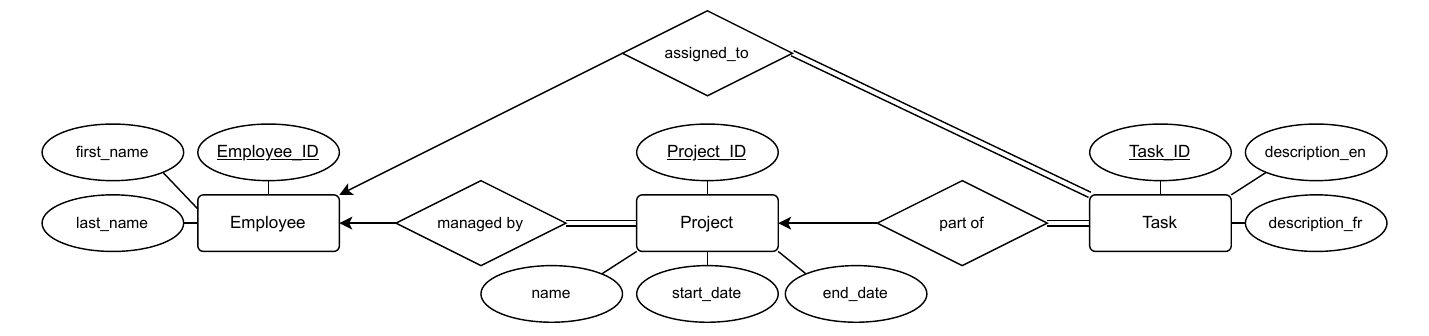}
    \caption{Entity-Relationship Diagram (ERD) representing the UoD of the data used in the protocol.}
    \label{fig:erd}
\end{figure}

Participants will be requested to complete five mapping tasks, some of which are interdependent. We can observe that different studies adopt different domains, some of which are specific (e.g., health care). We argue that participants should question the domain used in the experiment. Therefore, we propose a domain that is sufficiently generic and accessible for participants to understand. Our protocol proposes mapping data about projects, project tasks, and employees who manage projects and are assigned tasks. Figure \ref{fig:erd} depicts the Universe of Discourse (UoD) of the data to be transformed into RDF. We use an Entity-Relationship Diagram (ERD), but JSON and XML files can easily represent the data.\footnote{Examples are provided in the GitHub repository under the ``assets'' directory.} To ensure attribute names do not confuse participants, we ensured all attributes are unambiguous. In this simple UoD, all relations are many-to-one, though this can be easily extended to many-to-many when transforming documents.

The tasks can be summarized as follows: 

\begin{enumerate}
    \item Generate instances of \texttt{ex:Employee} with their first and last-names. The IRIs of employees are based on the name.
    \item Generate instances of \texttt{ex:Project} with their name, start- and end-date. Both dates are of the type \texttt{xsd:date}, allowing us to assess the creation of typed literals. The IRIs of projects are based on the project's ID.
    \item Generate \texttt{ex:managedBy} properties from projects to employees.
    \item Generate instances of \texttt{ex:Task} with their descriptions (in two languages). The IRIs of tasks are based on the task's ID. The descriptions allow us to assess the creation of language tags.
    \item Generate \texttt{ex:of} and \texttt{ex:assignedTo} properties from tasks to, resp., projects and employees.  
\end{enumerate}

We point out that the mappings mainly focus on RML-core \cite{DBLP:conf/icwe/Iglesias-Molina23} functionality. Part of RML-core are multi-valued expression maps, which are irrelevant for CSV files. People can easily adapt the JSON files to provide one or more task descriptions in different languages to test more complex language maps, for instance. 

We draw attention to the fact that employees' IRIs are based on their names, whereas the project data sources refer to employees via their IDs. This requires users to join the data in the two sources. In the case of RML, this requires users to ``join'' the two sources either at the level of the logical source or by using a referencing object map.

\subsection{Variants}
\label{Variants}

We stated that the tasks focus on RML-core functionality, which is also covered by other KGC languages and techniques such as ShExML \cite{DBLP:journals/peerj-cs/Garcia-Gonzalez20} and SPARQL Anything \cite{DBLP:journals/corr/abs-2106-02361}. While one can argue that the set of desired functionalities is limited, \cite{DBLP:conf/icwe/Iglesias-Molina23} covered the requirements of all RML extensions. Formulating tasks that cover practically all possible features and scenarios, not only in time but also in data source complexity, is not feasible. However, when a particular feature or scenario needs to be investigated, the protocol can be amended. This section describes how this protocol can be adapted and used to compare different tools, features, or aspects.

\begin{itemize}
    \item One can provide the same tasks to two different groups when comparing languages, techniques, or tools. One can compare different mapping languages (e.g., ShExML vs. RML), compare editors vs. ``bare bones''  languages (e.g., RMLEditor vs. RML), compare languages and abstractions of languages (e.g., RML vs. YARRRML), and even different editors and languages.
    \item When the aim is to compare a tool or language's support for an advanced KGC requirement such as named graphs, collections and containers, or RDF-star (among others), then one can take this protocol as is for one group, and only change the last two tasks for the second in which those requirements are covered for the second group. The first three tasks, which are shared, provide a basis for comparing the two groups. This design ensures that both groups share a comparable foundation (tasks 1-3), allowing us to isolate and evaluate the impact of the new features introduced in the final tasks.
    \item We have proposed a generic and accessible domain for the protocol to ensure broad applicability. However, the protocol can be adapted to include similar tasks within a different (and domain-specific) context. This adaptation would enable researchers to assess whether a language or tool designed for a particular application domain performs better in its setting. In such cases, it is important to compare the tool or language across both the original (generic) domain and the adapted (domain-specific) version.
\end{itemize}

Participants are assumed to have access to prepared ``resources'' or ``environments'' to focus on the tasks. In the case of RML, for instance, the logical sources would be provided in the tool or for them to copy and paste. This allows researchers to assess the languages and tools with respect to these aspects by giving one group the prepared artifacts and requesting the other to formulate the logical sources themselves. We deem this a special case of comparing a baseline with an extension as described above, but where the five tasks remain unchanged.

\subsection{Post-questionnaire}
Both SUS and PSSUQ are used to measure usability. SUS is adequate for a rapid and general measure of a system's usability. Still, the latter offers more advantages because it assesses three aspects of a system: system usefulness, information quality, and interface quality. Furthermore, there is a question about the system as a whole, which allows one to dampen the perception of individual aspects. The original PSSUQ survey uses 19 questions (as adopted by \cite{DBLP:conf/esws/JuniorDO17}, for instance), but recent iterations have removed three redundant questions.

As for the perceived (mental) workload, we adopt both the Workload Profile (WP) \cite{tsang1996diagnosticity} and the NASA Task Load Index (NASA-TLX) \cite{hart2006nasa}.

\begin{itemize}
    \item WP adopts a theory in which participants have different capacities (dimensions) related to the stage, mode, input, and output of information processing. The eight dimensions are each quantified through subjective rates, and participants must rate the proportion of attentional resources used for performing a given task with a value from 0 to 100 after task completion. A rating of 0 means that the task placed no demand, while 100 indicates that it required maximum attention. The WP of a participant is calculated as $\frac{1}{8}\sum_{i=1}^{8}{d_i}$.
    \item NASA-TLX has been validated in several domains \cite{hart2006nasa} and combines six factors believed to influence the mental workload. Each factor is quantified with a subjective judgment and a weight computed via a paired comparison procedure.  For each possible pair of the six factors, participants must decide which factor contributed the most to the mental workload during the task. The weights $w$ are the number of times each dimension was selected. The possible weights range from 0 (irrelevant) to 5 (most important). The final score is computed as a weighted average, considering the subjective rating of each attribute $d_i$ and the corresponding weights $w_i$: $\frac{1}{15}\sum_{i=1}^{6}{d_i*w_i}$. It is possible to calculate the scores by eliminating the weighted procedure, which yields the so-called \textit{Raw TLX}.
\end{itemize}

Both instruments are used in industry and research. You may notice that both instruments use different rating systems, which may confuse participants in a paper survey. Erroneous inputs can be prevented by adopting electronic forms. We choose not to harmonize the scales, which has been done in \cite{DBLP:conf/esws/JuniorDO17}, for instance, to obtain results that can be compared to other studies faithfully adopting those instruments.

Studies should explicitly report the method used to calculate performance measures, such as accuracy. For instance: 

\begin{itemize}
    \item Accuracy should be determined by (1) \textit{graph isomorphism} (did they generate the expected graph, which is true or false), and, for more nuanced numbers, (2) \textit{precision} (the proportion of triples that are generated that are in the expected graph), (3) \textit{recall} (the proportion of expected triples that are in the generated graph), and (4) \textit{F-measure} combining precision and recall.
\end{itemize}

This approach accounts for situations where a participant generates additional triples, for instance. Figures should be reported both on the task and global levels. 

\begin{itemize}
    \item Efficiency is often erroneously conflated with \textit{task execution time}. The former is a broad concept that measures how well resources are used, which is not only time. Most studies measured the time it took for tasks to be completed. We recommend studies to report on task execution time, and the method to measure it. One can manually record time or use software to record user interactions to time the tasks. Another approach is to request users to report on the time or use electronic forms that keep track of time.
\end{itemize}

While we strongly encourage placing a time limit on the tasks for the experiment to obtain comparable results, there are two important cases to track: \textit{did not finish} and \textit{did not start}. The former may indicate insufficient time left to finish a task or that the task was too difficult. The latter merely indicates that the user never started the task. 

We propose to limit the protocol to these five metrics (four on accuracy and one related to task execution time). Studies are free to include other metrics, such as the number of times a mapping was executed (i.e., trial and error), but that would indirectly impact the task execution time. 

As the experiment should not be too time-consuming, we avoided interviews to obtain qualitative feedback. We also avoid ``think aloud'' experiments as they can impact the cognitive load. Whether a study reports on it or not, we recommend studies to report on any additional instruments they used. There is, however, a qualitative dimension to our protocol, as the PSSUQ does leave room for comments on each of the 16 questions. 

\section{Results and Analysis}
\label{Results and Analysis}

As part of our protocol, we recommend a structured approach for reporting collected data. As mentioned, while many user studies focus primarily on presenting averages and standard deviations, we emphasize the importance of extending these reports to include statistical tests. This ensures robust comparisons between groups and tools, enhancing the general reliability and interpretability of the experiments. The following describes the recommended aspects and tests to be considered when reporting results and analysis.

\begin{description}
    \item[Reliability and Internal Consistency] Reliability refers to the degree to which the items within a test or survey consistently measure the same construct. High internal consistency strengthens the statistical reliability of metrics, thereby enhancing the validity of group comparisons. We recommend using \textit{Cronbach's Alpha} to evaluate internal consistency. Higher alpha coefficients indicate greater shared covariance among items, suggesting they assess the same underlying concept. A Cronbach's Alpha value of $\geq 0.7$ is generally considered acceptable. \cite{peterson1994meta}
    \item[Data Normality] Data normality refers to how much data distribution aligns with a normal curve. While normality is not always required, t-tests and ANOVA assume that data follows such a distribution. ANOVA is relatively robust when data is not normally distributed, when sample sizes are large, but that is difficult when dealing with user studies. We, therefore, require studies to test for normality and report on normality. Participants may, if they wish, use other statistical measures. As the sample sizes of a group will likely not exceed 50, we propose the \textit{Shapiro-Wilk test} to assess normality. In this test, the sample is compared to a theoretical normal distribution.
    \item[Homogeneity] Some statistical tests, such as ANOVA, assume that the variances across groups are equal. This is known as the \textit{homogeneity of variances}. Again, this assumption should be verified as part of the analysis. \textit{Levene's Test} is a standard method for evaluating this assumption.
    \item[Group Comparisons] To determine whether differences between groups are statistically significant, researchers should employ well-established statistical tests. The choice of test depends on the assumptions about the data. The recommended tests when the data is normally distributed (also known as parametric tests) are \textit{Welch's t-test} when comparing two groups and \textit{ANOVA} when comparing more than two groups simultaneously. Both tests assume normality and homogeneity of variance. The recommended tests when the data is not normally distributed (also known as non-parametric tests) are the \textit{Wilcoxon test} for comparing two groups and the \textit{Kruskal-Wallis test} for multiple.
    \item[Correlation Analysis] Correlation methods assess the strength and direction of the relationship between variables, which can provide deeper insights into study outcomes. For instance, examining correlations between usability, accuracy, and mental workload can reveal relationships in user behavior. When data is normally distributed, we recommend the \textit{Pearson's Correlation} to measure the strength of a relationship, for example, between accuracy and usability. Otherwise, one should use the \textit{Spearman's Correlation}. Researchers must report on correlations between all relevant variables (e.g., usability and mental workload, usability and accuracy, etc.) to provide a comprehensive analysis.
    \item[Transparency and Accessibility] To promote transparency and reproducibility, the data and the statistical tests performed should be publicly available online, provided that all personally identifiable information is removed and participant anonymity is preserved in accordance with ethical research standards.
    
    Access to data, analysis scripts, and detailed methodology facilitates validation and enhances the study's credibility. Moreover, sharing such data would allow one to compare results across studies more easily, provided the conditions are similar.
\end{description}


\section{Discussion}
\label{Discussion}

We presented a comparison of user studies in the KGC domain, and we noticed that all studies were different. This makes it impossible to compare KGC languages, tools, and software. To this end, we analyzed the related work, distilled elements we appreciated, and proposed others to establish a common protocol. As such, we proposed a resource, a user study protocol, that provides the KGC community with a better way to present, compare, and scrutinize contributions. 

When designing the protocol, we selected and refined elements from the state-of-the-art that we appreciated. Examples include the use of accuracy and task execution time as simple measures, the use of PSSUQ over SUS to obtain more fine-grained information on usability, usefulness, and information quality, and measuring the mental workload right after the task. Reporting on the correlations between the perceived usability, task execution, and mental workload could shed interesting insights. \cite{DBLP:conf/hworkload/JuniorDLO18} We provided guidelines on what statistical techniques to use when reporting on user studies with this protocol, as most merely reported on averages.

Relatively novel compared to the related work is our informed decision to adopt an accessible domain, focus on RML-core functionality as a basis, and formulate five tasks. In Section \ref{Proposed User Study Protocol}, we provided a rationale that, when comparing two groups, the last two tasks can be replaced for one group so that extensions or variants within the same language or tool can be compared. As such, the resource is sufficiently general for use in the KGC community, and the approach in designing this protocol may inspire others within the Semantic Web community. 

The resources have not only been made available with a DOI\footnote{\url{https://zenodo.org/records/14507455}} on a long-term preservation platform, but they are also available on a GitHub repository. The latter allows peers to contribute to the project. The directory structure we use allows for variants of the protocol to be made available.

The protocol has not yet been used at this stage, but its first use is planned for the spring of 2025. However, many of its separate elements are drawn from existing studies. As such, those parts have already been validated in the community. We also aim to engage with the wider KGC community via the W3C working group on adopting this protocol across different institutions. 

\section{Conclusions}
\label{Conclusions}

Prior KGC user studies used different protocols, making comparison impossible.  
This paper thus highlights the lack of standardized protocols in user studies related to KGC, making it difficult to compare different studies. We present a new protocol to address these inconsistencies, focusing on participant selection, task design, and evaluation metrics. The protocol suggests detailed guidelines for recruiting participants and disclosing potential biases. The process guidelines for informed consent, pre-questionnaires, familiarization activities, task execution, and post-questionnaires. Five specific mapping tasks are proposed, which can be solved with the equivalent of the RML-Core specification. The protocol recommends using the Post-Study System Usability Questionnaire (PSSUQ) for usability, and the NASA Task Load Index (NASA-TLX) for mental workload, among others. We designed the protocol in such a way that one can analyze one group, or compare groups. To this end, we provided guidelines on which statistical instruments to use.  

This protocol aims to provide a more comparable evaluation of KGC user studies, ultimately leading to more effective tools for knowledge graph construction. As such, the protocol is an essential artifact for future longitudinal and comparative studies. We recognize that the proposal has been constructed in a bottom-up fashion for this community, and future work should look into aligning our proposal with methods for comparing the usability of different (information) systems, such as \cite{10.1007/978-3-030-45002-1_39}.  Finally, future work also involves encouraging the adoption of this protocol by various KGC scholars. While ambitious, it is hoped that this protocol will form the basis of a new, open repository of KGC user studies. 

\begin{acknowledgments}
We thank the reviewers for their many (many) thoughtful comments, which greatly improved the paper. Their feedback was invaluable.
\end{acknowledgments}

\section*{Declaration on Generative AI}
During the preparation of this work, the author(s) used Grammarly to improve grammar, check spelling, and reword. After using these tool(s)/service(s), the author(s) reviewed and edited the content as needed and take(s) full responsibility for the publication's content.

\end{document}